\pdfoutput=1

\documentclass[twocolumn,aps,prl,showpacs,superscriptaddress]{revtex4}

\usepackage{graphicx}
\usepackage{amsmath,amssymb}
\usepackage{color}
\usepackage[colorlinks=true,citecolor=blue,urlcolor=blue]{hyperref}

\def\ket#1{|\,#1\,\rangle}

\begin{document}


\title{Experimental Observation of Hardy-Like Quantum Contextuality}



\author{Breno Marques}
\affiliation{Department of Physics, Stockholm University, S-10691 Stockholm, Sweden}

\author{Johan Ahrens}
\affiliation{Department of Physics, Stockholm University, S-10691 Stockholm, Sweden}

\author{Mohamed Nawareg}
\affiliation{Department of Physics, Stockholm University, S-10691 Stockholm, Sweden}
\affiliation{Instytut Fizyki Teoretycznej i Astrofizyki, Uniwersytet
Gda\'{n}ski, PL-80-952 Gda\'{n}sk, Poland}

\author{Ad\'an Cabello}
\email{adan@us.es}
\affiliation{Departamento de F\'{\i}sica Aplicada II, Universidad de Sevilla, E-41012 Sevilla, Spain}

\author{Mohamed Bourennane}
\email{boure@fysik.su.se}
\affiliation{Department of Physics, Stockholm University, S-10691, Stockholm, Sweden}


\date{\today}


\begin{abstract}
 Contextuality is a fundamental property of quantum theory and a critical resource for quantum computation. Here, we experimentally observe the arguably cleanest form of contextuality in quantum theory [A. Cabello \emph{et al.}, Phys. Rev. Lett. \textbf{111}, 180404 (2013)] by implementing a novel method for performing two sequential measurements on heralded photons. This method opens the door to a variety of fundamental experiments and applications.
\end{abstract}



\maketitle


{\em Introduction.---}The discovery that quantum probabilities cannot be reproduced by a joint probability distribution over a single probability space \cite{Bell66, KS67} implies that, in quantum theory (QT), measurement outcomes cannot be preassigned independently of the measurement ``context'' (i.e., of the set of other compatible measurements that may be carried out). Consequently, some quantum predictions of QT cannot be reproduced by any noncontextual hidden variable (NCHV) theory. In this sense, it is said that QT exhibits contextuality.

Recently, contextuality has been identified as a critical resource for universal quantum computation via ``magic state'' distillation \cite{HWVE13, DGBR14} and for measurement-based quantum computation \cite{Raussendorf13}. Contextuality is also the underlying property behind nonlocality \cite{Bell64} and its applications, e.g., cryptography \cite{Ekert91}, reduction of communication complexity \cite{BZPZ04}, and randomness expansion \cite{PAMBMMOHLMM10}.

All this makes the following question of fundamental importance: What is the simplest form of contextuality and how can it be observed? It has been recently pointed out \cite{CBTB13} that there is a form of contextuality that is analogous to ``the simplest and cleanest'' \cite{Mermin95} form of nonlocality found by Hardy \cite{Hardy92,Hardy93}. In this Letter we present the first experimental observation of this form of contextuality.


{\em Contextuality made simple.---}The result in Ref.~\cite{CBTB13} can be summarized as follows. Consider five imaginary boxes, numbered from 1 to 5, which can be either full or empty. $P_{\ket{\psi}}(0,1|i,j)$ denotes the joint probability in state $|\psi\rangle$ that box $i$ is empty and box $j$ is full. Suppose that
\begin{subequations}
 \label{uno}
\begin{align}
 &P_{\ket{\psi}}(0,1|1,2)+P_{\ket{\psi}}(0,1|2,3)=1, \label{unoa}\\
 &P_{\ket{\psi}}(0,1|3,4)+P_{\ket{\psi}}(0,1|4,5)=1. \label{unob}
\end{align}
\end{subequations}
Then, assuming that the outcomes are noncontextual, one would lead to the conclusion that
\begin{equation}
 P_{\ket{\psi}}(0,1|5,1)=0.
 \label{dos}
\end{equation}
However, in QT conditions (\ref{unoa}) and (\ref{unob}) occur while prediction (\ref{dos}) fails. Instead of Eq.~(\ref{dos}), QT predicts
\begin{equation}
 P_{\ket{\psi}}(0,1|5,1)=\frac{1}{9}.
 \label{qdos}
\end{equation}


{\em Why Hardy-like experiments are difficult.---}As pointed out by Mermin, although Hardy-like proofs ``reign supreme in the {\em gedanken} realm,'' they ``provide a rather weak basis for a laboratory violation of the experimentally relevant inequality'' \cite{Mermin94a}. Each of these proofs is equivalent to a violation of an inequality: the Clauser-Horne-Shimony-Holt Bell inequality \cite{CHSH69} in the case of Hardy's proof of nonlocality \cite{Mermin94a}, and the Klyachko-Can-Binicio\u{g}lu-Shumovsky (KCBS) noncontextuality (NC) inequality \cite{KCBS08} in the case of the Hardy-like proof of contextuality \cite{CBTB13}. The difficulty comes from the fact that the violation is small compared to the violation that can be achieved when the constraints of a Hardy-like proof are removed. Consequently, the experimental observation of Hardy-like nonlocality or contextuality requires very precise state preparation and measurements and is much more difficult than observing a violation of a Bell or NC inequality.

Despite these difficulties, several experiments have tested Hardy's nonlocality \cite{TBMM95,BBDH97}. An equivalent experiment for contextuality faces an additional obstacle: it cannot be implemented by measuring different subsystems of a composite system, as in Bell-inequality experiments, or different degrees of freedom of a single system \cite{MWZ00,HLBBR03}, but requires sequential measurements on the same system. Moreover, (1) the sequential measurements must be compatible \cite{GKCLKZGR10} and (2) every measurement must be carried out using the same device in any context \cite{ABBCGKLW13}. In addition, (i) the probabilities in Eqs.~(\ref{unoa}) and (\ref{unob}) must sum up to $1$ within the experimental error, (ii) the probability in Eq.~(\ref{qdos}) must be in agreement with the quantum prediction, and (iii) when considered together, the probabilities must violate the KCBS inequality.


\begin{figure}[tb]
\begin{center}
\centerline{\includegraphics[width=1\columnwidth]{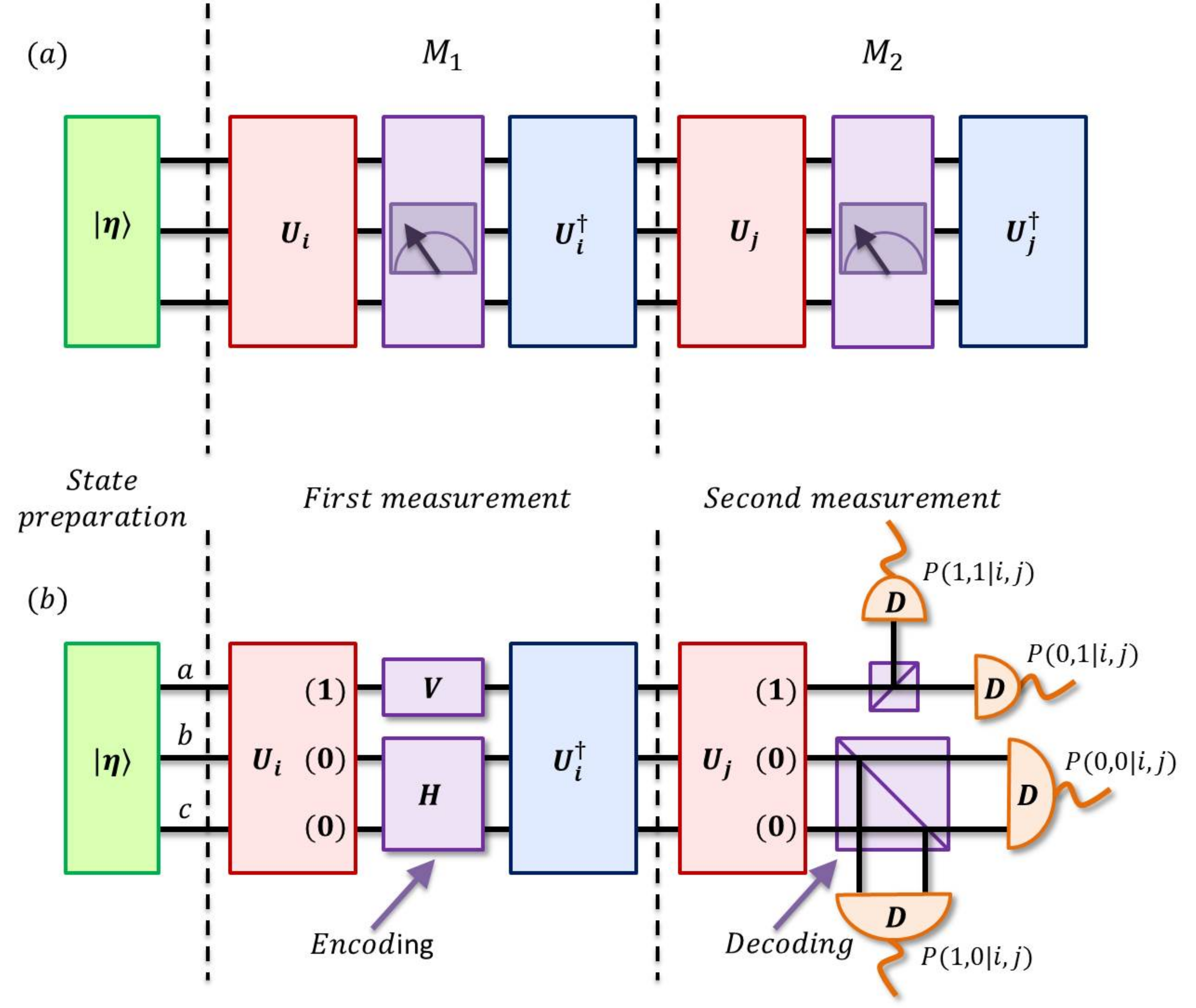}}
\caption{(color online) Sequential measurements on the same photon. (a) The system is prepared in state $\ket{\eta}$ and then submitted to two nondemolition measurements, one after the other. The measurement $i$ is implemented by means of a unitary operation $U_i$ that maps the eigenstate of $i$ with eigenvalue $1$ into the desired path, followed by an operation that stores the outcome in a detector, followed by the inverse unitary operation $U_i^\dag$ that rotates to the basis used in the state preparation. The measurement of $j$ is implemented similarly. (b) Our photonic implementation. The initial state and the measurements refer to the photon's path degrees of freedom. The outcome of $i$ is encoded in the photon's polarization. Then the unitary operation $U_j$ corresponding to $j$ is applied. Then, the outcome of $i$ is decoded using polarizing beam splitters. The outcomes of $i$ and $j$ are given by the detector that clicks (since each detector corresponds to a combination of outcomes). There is no need to implement $U_j^\dag$, since no additional measurements will be performed.}\label{Fig01}
\end{center}
\end{figure}

\begin{figure}[bt]
\begin{center}
\centerline{\includegraphics[width=0.9\columnwidth]{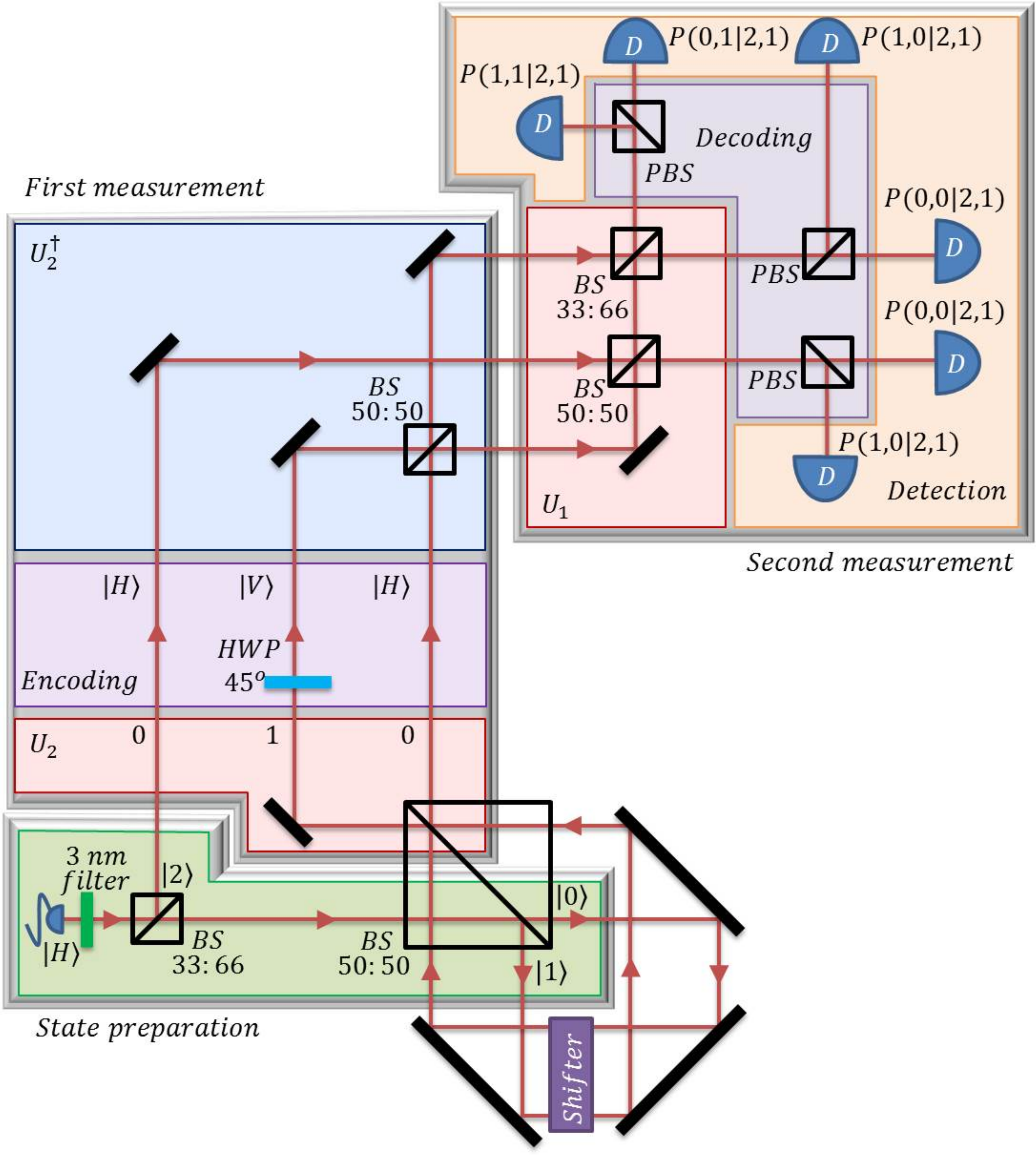}}
\caption{(color online) Experimental setup for measuring observable $2$ (in the first place) and observable $1$, given by Eq.~(\ref{vec}), on the initial state (\ref{state}). After preparing state (\ref{state}), the paths corresponding to $\ket{0}$ and $\ket{1}$ are injected into a Sagnac interferometer to control the phase difference. The phase shifter consists of two quarter wave plates and one HWP in between them. The result of the unitary operation $U_2$ is given by the Sagnac outputs together with the third path. Then the output of observable $2$ is encoded by rotating the polarization in one of the paths with the help of a HWP at $45^{\circ}$. Then $U_2^\dagger$ is applied to go back to the basis used in the state preparation, and the unitary operation $U_1$ is implemented in a similar way. The outcome of observable $2$ is finally decoded using polarizing beam splitters (PBSs), so the single photon detector that clicks gives the outcomes of observables $2$ and $1$. All coincidence counts between the signal and idler photons are registered using an eight-channel coincidence logic with a time window of $1.7$~ns. The number of detected photons was approximately $2~\times~10^3$ per second and the total time used for each experimental configuration was $10$~s.}
\label{Fig02}
\end{center}
\end{figure}

{\em Experimental setup.---}Our experiment adopts a novel method for performing two sequential measurements on the path degrees of freedom of the same photon. See Fig.~\ref{Fig01}. In our experiment, the physical systems are defined by single photons in a three-path setup. The basis vectors $\ket{0}$, $\ket{1}$, and $\ket{2}$ correspond to finding the photon in path $a$, $b$, or $c$, respectively. Each run of the experiment consists of preparing a single photon in a given state and measuring two compatible observables, $i$ and $i+1$ (or $i+1$ and $i$), sequentially. Single photons are generated from a heralded single photon source through a spontaneous parametric down-conversion process. The idler photon is used as the trigger. The initial state of the signal photon in the path degrees of freedom is
\begin{equation}
\label{state}
 |\eta\rangle = \frac{1}{\sqrt{3}}(1,1,1)^T,
\end{equation}
where $T$ means transposition. This state is prepared by combining two beam splitters (BSs), the first with a reflectivity-to-transmitivity ratio of 33:66 and the second with a 50:50 ratio. See Fig.~\ref{Fig02}. To exactly define the spatial and spectral properties of the signal photon, the source is coupled into a single mode fiber and passed through a narrow-band interference filter ($F$).

To perform two sequential measurements on the same photon we implement the scheme shown in Fig.~\ref{Fig01} (b) in which the outcome of the first measurement (that only addresses the spatial degrees of freedom) is encoded in the polarization of the photon before the second measurement (that also only addresses the spatial degrees of freedom). For that, the signal photon is initially horizontally polarized; thus encoding the outcome of the first measurement simply requires rotating the polarization in one of the paths after the unitary transformation corresponding to the first measurement.

Two measurements $\mu_1$ and $\mu_2$ are compatible if there is a measurement $\mu$ such that the outcome set of $\mu$ is the Cartesian product of the outcome sets of $\mu_1$ and $\mu_2$, and, for all states, the outcome probability distributions for $\mu_1$ or $\mu_2$ are recovered as marginals of the outcome probability distribution of $\mu$. In our experiment, we took advantage of the fact that $\mu_1$ and $\mu_2$ are sharp quantum measurements on path degrees of freedom of a photon. In this case, there is an algorithm for constructing the corresponding measurement devices for $\mu_1$ and $\mu_2$ \cite{RZBB94}. Then, measuring $\mu$ is equivalent to measuring sequentially $\mu_1$ and $\mu_2$. Compatibility can then be tested by checking that the order of $\mu_1$ and $\mu_2$ does not affect the probabilities. Perfect compatibility is only limited by our ability to construct devices corresponding to the exact unitary transformations needed and by imperfections when combining them.

\begin{figure}[tb]
\begin{center}
\centerline{\includegraphics[width=0.66\columnwidth]{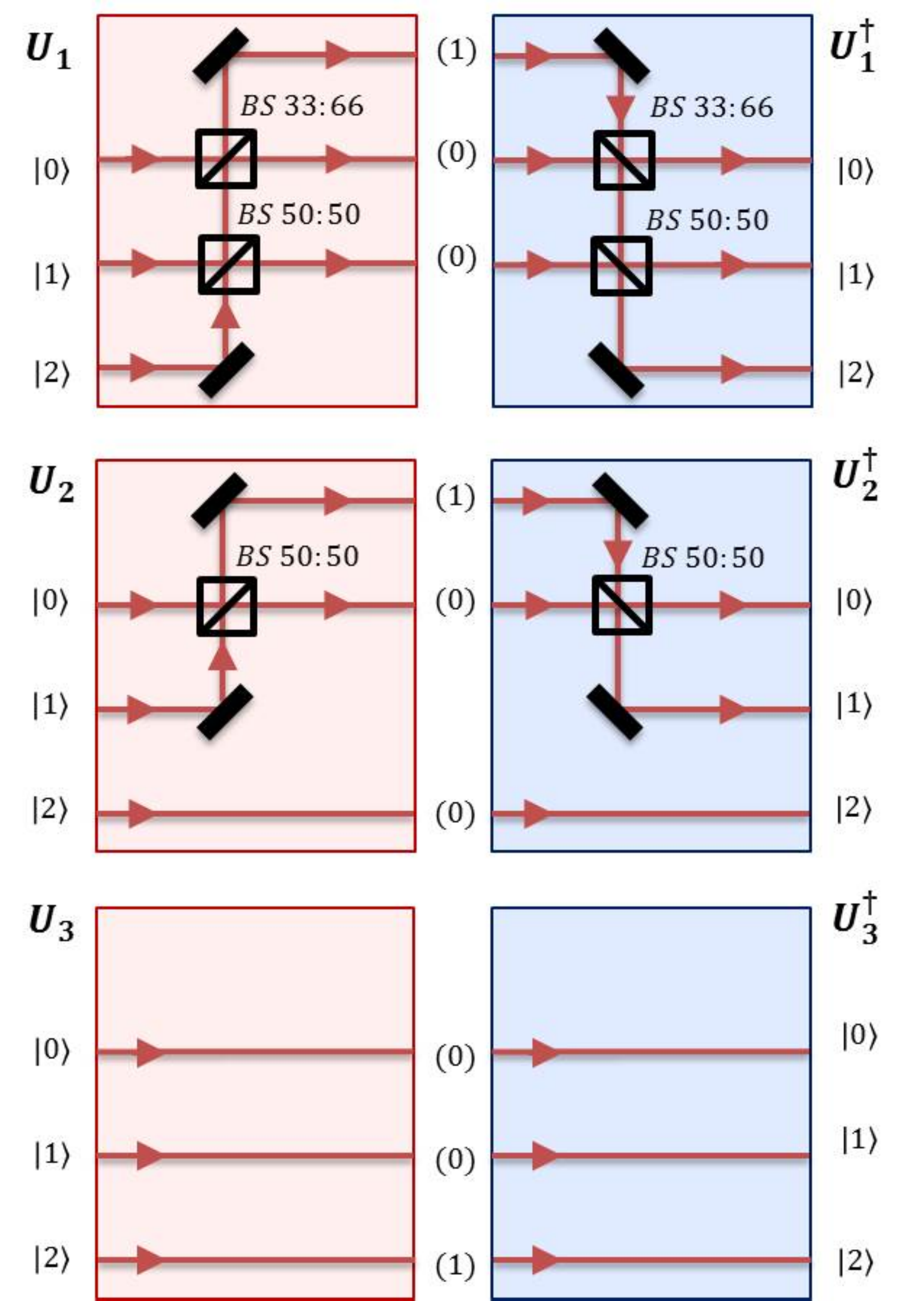}}
\caption{(color online) Experimental setups for the unitary operations $U_i$ (in red) and $U_i^\dag$ (in blue) for $i=1,2,3$. They consist of combinations of BSs with 33:66 and 50:50 splitting ratios. The setups for $U_4$ and $U_4^\dag$ are the same as those for $U_3$ and $U_3^\dag$, respectively, but with the following relabeling of paths: $\ket{0}_{U3}\rightarrow\ket{1}_{U4}$, $\ket{1}_{U3}\rightarrow\ket{2}_{U4}$, and $\ket{2}_{U3}\rightarrow\ket{0}_{U4}$. The setups for $U_5$ and $U_5^\dag$ are the same as those for $U_2$ and $U_2^\dag$, respectively, but with the following relabeling of paths: $\ket{0}_{U2}\rightarrow\ket{1}_{U5}$, $\ket{1}_{U2}\rightarrow\ket{2}_{U5}$, and $\ket{2}_{U2}\rightarrow\ket{0}_{U5}$.}\label{Fig03}
\end{center}
\end{figure}

The measurements $i$ in our experiment, with $i=1,\ldots,5$, are those represented by the projectors
$|v_i\rangle\langle v_i|$ on the following states:
\begin{subequations}
\label{vec}
\begin{align}
 &|v_1\rangle = \frac{1}{\sqrt{3}}(1,-1,1)^T,\\
 &|v_2\rangle = \frac{1}{\sqrt{2}}(1,1,0)^T,\\
 &|v_3\rangle = (0,0,1)^T,\\
 &|v_4\rangle = (1,0,0)^T,\\
 &|v_5\rangle = \frac{1}{\sqrt{2}}(0,1,1)^T.
\end{align}
\end{subequations}
The possible outcomes are 1 and 0. Each measurement $i$ consists of a unitary transformation $U_i$ to project the qutrit onto the two eigenspaces of $i$, followed by a recording of the outcome. In our case, the outcome of the first measurement is encoded in the polarization of the photon by adding a half wave plate (HWP) in one of the paths. Then the inverse unitary transformation $U_i^\dagger$ is implemented in order to rotate back to the initial basis. Unitary transformations $U_i$ and $U_i^\dag$ with $i=1,\ldots,5$ were implemented by mapping $\ket{v_i}$, i.e., the quantum state corresponding to eigenvalue $1$, to path $a$ and mapping the subspace corresponding to eigenvalue $0$ to the remaining two paths $b$ and $c$. The devices for the unitary transformations for the five measurements $i$ are shown in Fig.~\ref{Fig03}. Figure~\ref{Fig02} shows the complete experimental setup corresponding to the sequential measurement of observables $2$ (in the first place) and $1$.


State (\ref{state}) and these measurements lead to the conditions needed for a Hardy-like proof of contextuality, namely,
\begin{subequations}
\begin{align}
 &P_{|\eta\rangle}(0,1|1,2)+P_{|\eta\rangle}(0,1|2,3)=\frac{2}{3}+\frac{1}{3},\\
 &P_{|\eta\rangle}(0,1|3,4)+P_{|\eta\rangle}(0,1|4,5)=\frac{1}{3}+\frac{2}{3},\\
 &P_{|\eta\rangle}(0,1|5,1)=\frac{1}{9}.
 \label{unoq}
\end{align}
\end{subequations}


{\em Experimental results.---}The experimental results are presented in Table~\ref{Table1}. The errors come from Poissonian counting statistics and systematic errors. The main sources of systematic errors are the slight imperfections in the optical interferometers due to nonperfect overlapping and intrinsic imperfections of the BSs and HWPs. The results are in very good agreement with the predictions of QT for an ideal experiment and are essentially insensitive to the order in which the measurements are performed. This contrasts with previous photonic experiments in which the independence of the order was not tested \cite{ARBC09,DHANBSC13}.

Taking the experimental results needed for the Hardy-like contextuality from Table~\ref{Table1}, we obtain
\begin{subequations}
\begin{align}
 &P_{\ket{\eta}}(0,1|1,2) + P_{\ket{\eta}}(0,1|2,3) = 0.981 \pm 0.021, \label{e1} \\
 &P_{\ket{\eta}}(0,1|3,4) + P_{\ket{\eta}}(0,1|4,5) = 0.987 \pm 0.012, \label{e2} \\
 &P_{\ket{\eta}}(0,1|5,1) = 0.110 \pm 0.005. \label{e3}
\end{align}
\end{subequations}


\begin{table}
\begin{center}
\begin{tabular}{c c c c c}
\hline \hline
$(i,i+1)$ & $P_{\ket{\eta}}(0,1|i,i+1)$ & $P_{\ket{\eta}}(1,0|i+1,i)$ & Ideal \\
\hline
$(1,2)$ & $0.635\pm0.020$ & $0.661\pm0.011$ & $0.667$ \\
$(2,3)$ & $0.332\pm0.008$ & $0.331\pm0.005$ & $0.333$ \\
$(3,4)$ & $0.330\pm0.004$ & $0.339\pm0.003$ & $0.333$ \\
$(4,5)$ & $0.650\pm0.008$ & $0.656\pm0.011$ & $0.667$ \\
$(5,1)$ & $0.111\pm0.003$ & $0.109\pm0.004$ & $0.111$ \\
\hline \hline
\end{tabular}
\end{center}
\caption{\label{Table1} Experimental results. The second and third column shows the probabilities when the measurements are performed in direct and reverse order, respectively. The fourth column shows the prediction of QT for an ideal experiment.}
\end{table}


These results were obtained by measuring $i$ in the first place in half of the runs and measuring $i+1$ in the first place in the other half. These results show a very good agreement with the predictions of QT for an ideal experiment and thus provide experimental evidence of the Hardy-like contextuality as shown by the following facts: (i) $P_{\ket{\eta}}(0,1|1,2) + P_{\ket{\eta}}(0,1|2,3)$ and $P_{\ket{\eta}}(0,1|3,4) + P_{\ket{\eta}}(0,1|4,5)$ are $1$ within the experimental error, (ii) $P_{\ket{\eta}}(0,1|5,1)$ is nonzero and in very good agreement with the value predicted by QT given by Eq.~(\ref{qdos}), (iii) the joint probabilities are almost independent of the order in which the measurements were performed, and (iv) the sum of the joint probabilities violates the KCBS inequality \cite{KCBS08}, namely,
\begin{equation}
 S= \sum_{i=1}^{5} P(0,1|i,i+1) \stackrel{\mbox{\tiny{ NCHV}}}{\leq} 2,
 \label{kcbs}
\end{equation}
where the sum is taken modulo $5$ and ``$\stackrel{\mbox{\tiny{ NCHV}}}{\leq} 2$'' indicates that $2$ is the maximum for NCHV theories. From the results in Eqs.~(\ref{e1})--(\ref{e3}), we obtain
\begin{equation}
 S_{\rm exp} = 2.078 \pm 0.038,
 \label{kcbs}
\end{equation}
in agreement with the quantum prediction for an ideal experiment.

Compatibility was enforced by choosing $i$ or $i+1$ to be represented by commuting operators. In addition, we checked compatibility in two ways. First, we checked that the probabilities do not depend on the order in which measurements were performed. Second, we counted the clicks in the detector corresponding to $(1,1|i,i+1)$. These detections would never occur in an ideal situation, since the eigenstates of $i$ and $i+1$ with eigenvalue $1$ are orthogonal. Our experiment was very close to this ideal situation, since these events only occurred with probabilities in the range $0.002 \pm 0.001$--$0.006 \pm 0.001$ for nine out of the ten configurations, and with probability $0.021 \pm 0.001$ for the most complex configuration, which is the one shown in Fig.~\ref{Fig02}.

The noncontextual upper bound of the KCBS inequality is derived under the assumption that the two measurements are perfectly compatible. However, experimental imperfections make this assumption only approximately satisfied. To show the significance of the experimental results in this case, we followed the approach used in previous experiments \cite{DHANBSC13}. It consists of assuming that the noncontextual upper bound of the inequality is valid for some fraction $(1-\epsilon)$ of the experimental runs, but must be corrected assuming the most adversarial scenario for the other fraction $\epsilon$. The parameter $\epsilon$ is defined as the average of $P(1,1|i,i+1)$ for the ten experimental configurations tested. In an ideal experiment $\epsilon$ would be zero. Our assumption is that the noncontextual upper bound of the KCBS inequality (namely $2$) is valid only for a fraction $1-\epsilon$ of the runs, while for the remaining fraction $\epsilon$ we assume the worst-case scenario in which the maximum of the KCBS inequality for general probabilistic theories (namely $\frac{5}{2}$) is reached. In our experiment, $\epsilon = 0.0062$. Therefore, the noncontextual upper bound of the KCBS inequality shifts from $2$ to $2 (1-\epsilon)+\frac{5}{2} \epsilon = 2.0031$. Nevertheless, the experimental value in Eq.~(\ref{kcbs}) still violates this bound. Notice, that this is also true if we define $\epsilon$ as the largest value of of $P(1,1|i,i+1)$ for the ten experimental configurations, since in this case the bound is shifted from $2$ to $2.0105$.


{\em Conclusion and further applications.---}By implementing a new method for performing two sequential measurements on the same photon, we have presented the first experimental observation of Hardy-like contextuality, which is arguably the conceptually cleanest form of contextuality in physics and connects the quantum violation of the simplest NC inequality \cite{KCBS08} with the Kochen-Specker theorem \cite{KS67}, thus providing the link between two fundamental results in QT \cite{CBTB13}.

In this experiment, we have observed for the first time with photons that the correlations between the outcomes of sequential measurements represented by commuting operators are independent of the order in which the measurements are performed. In addition, the experiment is precise enough to confirm the small quantum violation of the relevant NC inequality. This shows that this method can be used for a variety of pending fundamental experiments demanding high precision sequential measurements, e.g., contextuality-based nonlocality \cite{Cabello10, CABBB12}, almost-state-independent contextuality \cite{KK12, TPKK13}, and contextuality-nonlocality monogamy \cite{KCK14}.

Finally, this method for sequential measurements on photonic systems opens the door to applications in quantum information processing that require both sequential measurements and transmission of quantum information between spatially separated parties, e.g., contextuality-based cryptography \cite{CDNS11} and dimension witnessing \cite{GBCKL14}.


\begin{acknowledgments}
 We thank K. Blanchfield and P. Mataloni for discussions. This work was supported by the Swedish Research Council, Ideas Plus (Polish Ministry of Science and Higher Education Grant No.\ IdP2011 000361), CAPES (Brazil), Project No.\ FIS2011-29400 (MINECO, Spain) with FEDER funds, and the FQXi large grant project ``The Nature of Information in Sequential Quantum Measurements.'' M.\,N. is supported by the international PhD project Physics of Future Quantum-Based Information Technologies: Grant No.\ MPD/2009-3/4 from the Foundation for Polish Science. A.\,C.\ thanks NORDITA and M.\,B.\ for their hospitality at Stockholm University.\\
\end{acknowledgments}



\end{document}